# Strong band kinks in magic-thickness Yb films arising from interfacial electron-phonon coupling

Yi Wu[1], Yuan Fang[1], Shuyi Zhou[1], Peng Li[1], Zhongzheng Wu[1], Zhiguang Xiao[1], Xiaoxiong Wang[2], Chao Cao[3], Tai-Chang Chiang[4*], Yang Liu[1,5,6*]

[1]Center for Correlated Matter and Department of Physics, Zhejiang University, Hangzhou, P. R. China

[2]College of Science, Nanjing University of Science and Technology, Nanjing, P. R. China

[3]Department of Physics, Hangzhou Normal University, Hangzhou, P. R. China

[4]Department of Physics and Frederick Seitz Material Research Laboratory, University of Illinois at Urbana-Champaign, Urbana, IL 61822, USA

[5]Zhejiang Province Key Laboratory of Quantum Technology and Device, Zhejiang University, Hangzhou, P. R. China

[6]Collaborative Innovation Center of Advanced Microstructures, Nanjing University, Nanjing, P. R. China

*Corresponding authors: yangliuphys@zju.edu.cn, tcchiang@illinois.edu

# Abstract

Interfacial electron-phonon coupling in ultrathin films has attracted much interest; it can give rise to novel effects and phenomena, including enhanced superconductivity. Here we report an observation of strong kinks in the energy dispersions of quantum well states in ultrathin Yb films grown on graphite. These kinks, arising from interfacial electron-phonon coupling, are most prominent for films with a preferred ("magic") thickness of 4 monolayers, which are strained and hole doped by the substrate. The energy position of the kinks agrees well with the optical phonon energy of graphite, and the extracted electron-phonon coupling strength $\lambda$ shows a large subband dependence, with a maximum value up to 0.6. The kinks decay rapidly with increasing film thickness, consistent with its interfacial origin. The variation of $\lambda$ is correlated with evolution of the electronic wave function amplitudes at the interface. A Lifshitz transition occurs just beyond the magic thickness where the heavy Yb 5$d$ bands begin to populate right below the Fermi level.

Electron-phonon coupling (EPC) plays an important role in condensed matter physics; it can drive superconductivity (SC) and charge-density-wave formation [1]. While EPC effects in bulk materials have been studied extensively, much less is known about interfacial EPC. Because of abrupt changes of the crystal potential at an interface [2], EPC at the interface can be much stronger than that in the bulk, possibly leading to enhanced properties in ultrathin films. A notable example is single-layer FeSe grown on $SrTiO_3$(001), for which a SC transition temperature ($T_C$) up to 60 K has been reported [3,4,5,6,7,8]. This single-layer $T_C$ is almost an order of magnitude higher than the $T_C$ of ~8 K in bulk FeSe. This remarkable effect has been attributed to a large interfacial EPC [9, 10,11,12, 13], and it points to a promising avenue towards achieving high $T_C$'s in interface-engineered systems [14].

Experimental signatures of interfacial EPC in thin films have been identified by angle-resolved photoemission spectroscopy (ARPES) via shake-off bands due to bosonic excitations [9] or temperature-dependent broadening of quasiparticle bands [2]. Theoretically, when the substrate phonon energy is much smaller than the film conduction band width, interfacial EPC can also give rise to kinks in the film quasiparticle dispersions [15], similar to those observed in bulk crystals such as certain cuprates [16,17,18] or in simple metal surfaces [19,20]. However, to the best of our knowledge, no such interface-EPC-induced kinks for films have been documented in the literature thus far. Here we report prominent kinks for the quantum well states (QWSs) in ultrathin Yb films grown on graphite, which imply a strong interfacial EPC in this system. Our extracted coupling constant $\lambda$ in the Yb films is as high as 0.6, which far exceeds the Bardeen-Cooper-Schrieffer (BCS) threshold for SC. It is interesting to note that both Yb and graphite are non-superconducting and exhibit a weak EPC in the bulk [21].

Our work on Yb films is also motivated by earlier studies suggesting that 4*f* electrons might be involved in the Fermi surface (FS) of ultrathin Yb films at low temperatures [22], analogous to the celebrated α phase of Ce [23,24]. Our results for Yb films on graphite demonstrate that the Yb 4*f* electrons remain localized and do not contribute to the FS. A heavy electron band just below the Fermi level, with effective mass up to 19 $m_e$, is observed for thick films and is attributed to the Yb 5*d* states. Prior studies of Yb films on W(110) [25,26,27], while informative, did not provide the fine band dispersions near the Fermi level to address the issues of 4*f*-5*d* occupancy and interfacial EPC effects.

Yb films were grown by molecular beam epitaxy (MBE) in an Omicron Lab10 growth chamber. ARPES measurements were performed by transferring the sample under ultrahigh vacuum from the growth chamber to a dedicated ARPES chamber. ARPES measurements were performed at ~20 K, using a Scienta-Omicron VUV-5k helium lamp and DA-30(L) electron analyzer. Most of the ARPES data were taken using He-II photons (40.8 eV), with an energy (momentum) resolution of ~10 meV (~0.01 Å$^{-1}$) (see [28] for details).

ARPES spectra for Yb films at various coverages, in units of monolayers (MLs), are shown in Fig. 1(a). The substrate graphite has no occupied bands or features within the probed energy and momentum ranges. Approximately parabolic bands emerge as the coverage increases; these correspond to QWSs [29]. The energies of these QWSs are governed by the Bohr-Sommerfeld quantization condition [29]

$$2k_z(E)Nt + \varphi_s + \varphi_i = 2n\pi, \quad (1)$$

where $k_z(E)$ is the perpendicular momentum as a function of energy *E* in accordance with the bulk band structure, *N* is the film thickness in units of ML, *t* is the thickness of one ML, $\varphi_s$ ($\varphi_i$) is the phase shift at the surface (interface), and *n* is the quantum number of each QWS subband. As *N* increases, more subbands should emerge within the same energy range. However, the results for all films with coverages less than 4 ML show the same set of subbands, and the ARPES intensities of these bands increase with coverage. Evidently, the films form islands with a preferred, or magic, height of 4 ML, as reported for other film systems [30,31,32,33]. The growth behavior is illustrated schematically in Fig. 1(e). For coverages above 15 ML, the QWSs become densely populated and merge into a quasi-continuum.

The magic-thickness growth behavior is corroborated by reflection-high-energy-electron-diffraction (RHEED) measurements (Fig. 1(b-d)). The patterns show streaks characteristic of two-dimensional films. The substrate is a highly oriented pyrolytic graphite (HOPG), which is made of graphite crystallites (~10 μm) well aligned along the $z$ axis but randomly oriented within the $xy$ plane. The Yb films with a hexagonal crystal structure grow along the (0001) direction [34] and are similarly randomly oriented within the $xy$ plane. Since the observed electronic states of Yb (Fig. 1(a)) are mainly derived from the nearly isotropic Yb 6$s$ electrons, the in-plane orientational averaging does not affect significantly the observed QWS dispersions. Selected line cuts of the RHEED intensity shown in Fig. 1(c) reveal that the in-plane lattice constant of the magic-height islands at coverages less than 4 ML is ~5% larger than that of a thick bulklike film, e.g., at 10 ML. This expansion is much smaller than the 10% lattice mismatch between bulk Yb and graphite (Fig. 1(e)). The intensity of the (10) film peak (Fig. 1(d)) shows a pronounced slope change near 4-ML coverage, which is consistent with a switchover from magic-height island growth to layer-by-layer growth.

Calculated electronic structures for the magic thickness (4 ML) and bulklike phase (10 ML) based on the experimental in-plane lattice constants are shown in Fig. 2(a) and (b), together with the ARPES data at coverages of 3 and 10 ML. The 3- and 4-ML ARPES spectral shapes are nearly identical due to formation of magic-height islands (Fig. 1(a)), but the 4-ML data contains a small contamination from 5-ML emission due to an unavoidable, although small, roughness. The calculations, based on density functional theory (DFT), are not necessarily accurate (details in [28]), but the overall band shapes are similar to the experiment (comparing the top and bottom panels in Fig. 2(a,b)). The bands near the zone center at high binding energies are approximately parabolic; these bands are derived from the Yb 6$s$ states based on the calculations. A key difference between the magic-thickness and bulklike phases is the presence in the latter of an intense emission from a fairly flat band just below the Fermi level, whose effective mass can be up to ~19 $m_e$ (inset in Fig. 2(c)). DFT calculations, despite indicating a negative effective mass, suggest that this heavy band is derived from the Yb 5$d$ states. Thus, a Lifshitz transition involving 5$d$ occupancy separates the magic-thickness phase from the bulklike phase. Theoretically, the Yb-6$s$-derived QWSs are at higher energies for the magic-thickness phase than the bulklike phase because of the 5% strain in the former (comparing the middle panels in Fig. 2(a,b)). This trend is qualitatively consistent with the experiment (comparing the top panels in Fig. 2(a,b)), although the experimental 6$s$ QWSs for the magic-thickness phase are even higher in energy than the calculation. Part of this energy shift might be caused by electron transfer from the Yb film to the substrate, which can be inferred from the Yb 4$f$ states to be discussed below.

One important question is the role of the Yb 4$f$ electrons, especially in view of earlier studies of a temperature-dependent phase transition in ultrathin Yb films [22], possibly analogous to the α-γ transition in Ce [23]. ARPES data for the magic-thickness and bulklike phases over a wide energy range (Fig. 2(c)) reveal two sets of flat 4$f_{5/2}$ and 4$f_{7/2}$ bands with a spin-orbit splitting of ~1.2 eV. Per earlier studies of various Yb-based compounds, the set of 4$f$ bands at higher (lower) energies is assigned to the bulk (surface) 4$f$ component, as labeled in Fig. 2(c)) [35]. All of the 4$f$ bands are far away from the Fermi level, corresponding to an electronic configuration of $4f^{14}$ for both phases. Thus, valence fluctuations involving the $4f^{13}$ configuration are irrelevant in the present case. A $4f^{13}$ Yb ion is expected to exhibit characteristic multiple peaks between -6 and -12 eV [36], which are not observed here. The bulk $4f$ states for the magic thickness are at slightly higher energies than those for the bulklike phase. This is consistent with a charge transfer across the interface that moves the QWSs toward higher energies as discussed earlier.

Strong kinks in the QWS dispersions for the magic-thickness phase (Fig. 3(a)) suggest a very large interfacial EPC. Here, the QWS subbands are labeled in terms of $n = 0, 1, 2, \ldots$ with $k_z(E)$ in Eq. (1) defined with respect to the zone boundary [29]. The kink is particularly strong for $n = 0$ (the topmost subband), but becomes weaker for increasing $n$. Following the standard procedure of analyzing EPC in bulk single crystals [18], we extrapolate from the portions of the QWS dispersions at larger binding energies to extract the bare-band dispersions $\varepsilon_k$ as indicated in the middle panel in Fig. 3(a). The real part of the self energy, $Re \sum(E)$, is the difference between the experimental band $E(k)$ and the bare band $\varepsilon_k$ (right panel in Fig. 3(a)). The resulting $Re \sum(E)$'s from the analysis for the different $n$'s show broad peaks at about -0.2 eV. This energy does not correspond to any known bosonic/phononic excitations from Yb [37]; instead, it corresponds well to the optical phonons in graphite, implying that it originates from a strong interfacial EPC. The strength of the interfacial EPC $\lambda$ can be estimated from

$$\lambda = \frac{(\frac{d\varepsilon}{dk})_{bare}}{(\frac{dE}{dk})_{renormalized}} - 1 . \qquad (2)$$

It is 0.58, 0.38, and ~0 for $n = 0, 1$, and 2, respectively, for the magic phase.

The kink feature diminishes quickly with increasing film thickness, as expected for interfacial EPC. The overall strength of the EPC for each QWS is given by an integral involving the QWS wave function over the film thickness. As a result, the interfacial contribution to the net

EPC must decrease with increasing film thickness [2]. Results of a similar analysis as above for 10 and 20-ML films are shown in Fig. 3(b). The extracted values of $\lambda$ for the $n = 0$ QWS is ~0.22 and ~0 for $N = 10$ and 20 films, respectively. Note that $Re\sum(E)$ for the $N = 10$ film is also peaked at about -0.2 eV, consistent with interfacial EPC, but the kink is much weaker.

The decrease of $\lambda$ from a maximum value of ~0.6 at $(N, n) = (4, 0)$ for increasing $N$ and $n$ is somewhat similar to that observed for the QWSs of Ag film on Fe(001) [2]. In that analysis, $\lambda$ is proportional to the fourth power of the amplitude of the QWS wave function (or square of the probability density) at the interface under a rigid-ion approximation, due to the large potential gradient at the interface. For the $n = 0$ QWS in the present case, the probability density has a maximum at each Yb atomic layer (bottom two curves in Fig. 4(a)). As $N$ increases, the probability density at each atomic layer diminishes correspondingly. This dilution effect leads to a reduction of the probably density at the interface, and $\lambda$ should decay approximately as $1/N^2$ (for the $n = 0$ QWS), as observed experimentally. For a given $N$, increasing $n$ by one results in a reduction of the number of probability density maxima in the film by one, as illustrated by the top three curves in Fig. 4(a). Hence the probability density maximum next to the interface moves inward toward the middle of the film. The resulting decrease of the probability density at the interface should cause $\lambda$ to decrease. Thus, the qualitative trends of the dependence of $\lambda$ on $N$ and $n$ are well explained by this simple model. A quantitative analysis, however, is difficult and will require accurate wave functions across the interface [38].

Fig. 4(b) shows a zoom-in view of the (4,0) QWS dispersion near the Fermi level, where the prominent kink is determined to be at -0.18 eV. For comparison, the in-plane optical phonons in graphite have energies between 150 and 200 meV (Fig. 4(c)) [39,40]. A prior study of FeSe/SrTiO$_3$ suggests that interfacial EPC could be strongly peaked at $\mathbf{q}_\| = 0$ due to a large ratio between the

in- and out-of-plane dielectric constants in two-dimensional films [9,41]. The same argument could apply in the present case. The kink energy of -0.18 eV corresponds well to a weighted average of the in-plane optical phonons near the $\Gamma$ and A points (green dashed circles in Fig. 4(c)). No kinks arising from other lower-energy phonons could be identified, indicating that couplings to these phonons might be much weaker.

The large interfacial EPC in the magic-thickness phase could be connected to the strain and possible charge transfer at the Yb/graphite interface. This implies strong interfacial bonding, likely caused by the small electronegativity of Yb. Similar interfacial charge transfer and strain effect have been observed in FeSe/SrTiO$_3$ [3], where $\lambda$ decays rapidly with $N$ and becomes negligibly small beyond 2 unit cells [9]. In the Yb/graphite case, however, large interfacial EPC occurs for a magic thickness of 4 ML and decays more gradually with thickness, similar to the Ag/Fe(001) case [2]. The different thickness dependence of $\lambda$ may be related to electronic wave mixing or phonon propagation across the interface.

The above discussion leads to the conclusion that the very strong kinks observed for some of the Yb QWSs are caused by an interfacial EPC effect that involves coupling of electrons in the Yb film and the in-plane optical phonons in the graphite substrate. This cross-interface fermion-boson interaction is strongest for a magic thickness of 4 ML and for the $n = 0$ QWS, which has the largest interfacial weight of the wave function. The magic thickness marks the boundary of a Lifshitz transition with Yb 5$d$ occupancy at larger film thicknesses. The 4$f$ electrons apparently do not play an important role in these changes. All of these observations indicate that the Yb/graphite system is highly unusual with remarkable properties that are governed by multiple effects including quantum confinement, strain, electronic phase transition, and interfacial EPC.

This work is supported by the National Key R&D Program of the MOST of China (Grant No. 2016YFA0300203, 2017YFA0303100) and the National Science Foundation of China (No. 11674280). TCC acknowledges support from the US Department of Energy under Grant No. DE-FG02-07ER46383. XXW acknowledges support from the Fundamental Research Funds for the Central Universities (No. 30917011338).

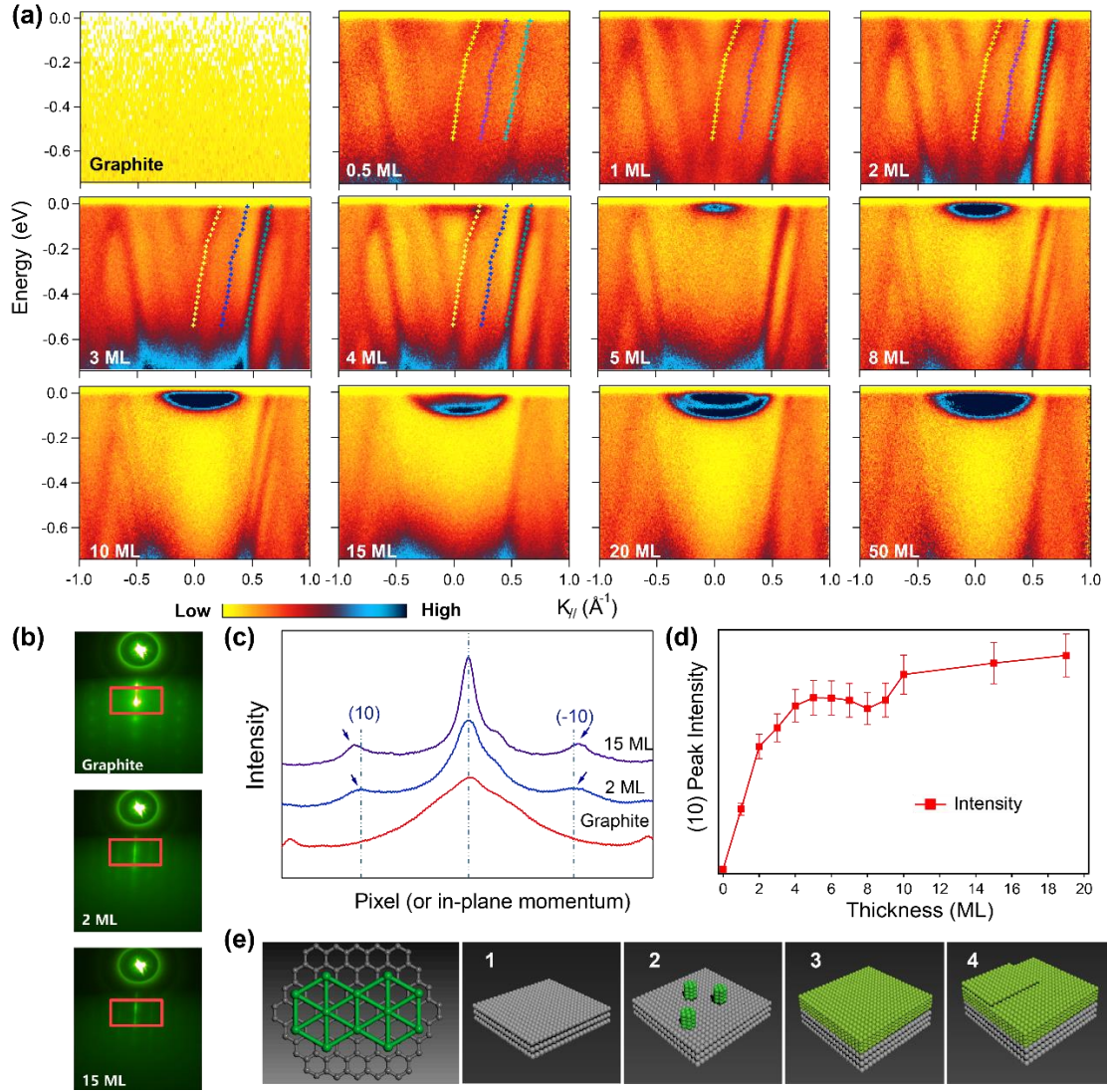

Fig. 1 (Color online). Film growth and formation of magic-height islands. (a) ARPES spectra of Yb films of various nominal coverages on graphite. The colored crosses are the extracted dispersions for the magic thickness 4 ML. (b) RHEED patterns taken at coverages of 0 ML (graphite), 2 ML, and 15 ML. (c) A line cut of RHEED intensity along the long side of the red boxes in (b) (integrated along the short side), demonstrating a change of in-plane lattice constant from 2 to 15 ML. (d) The (1,0) RHEED peak intensity as a function of film coverage shows a slope change near 4 ML. (e) Left: a perfect lattice match would require stretching the Yb lattice by ~10%. Right: cartoons illustrating growth steps (1-4). Below 4 ML, magic islands with 4 ML height are formed, and this is followed by quasi layer-by-layer growth above 4 ML (possibly with roughness).

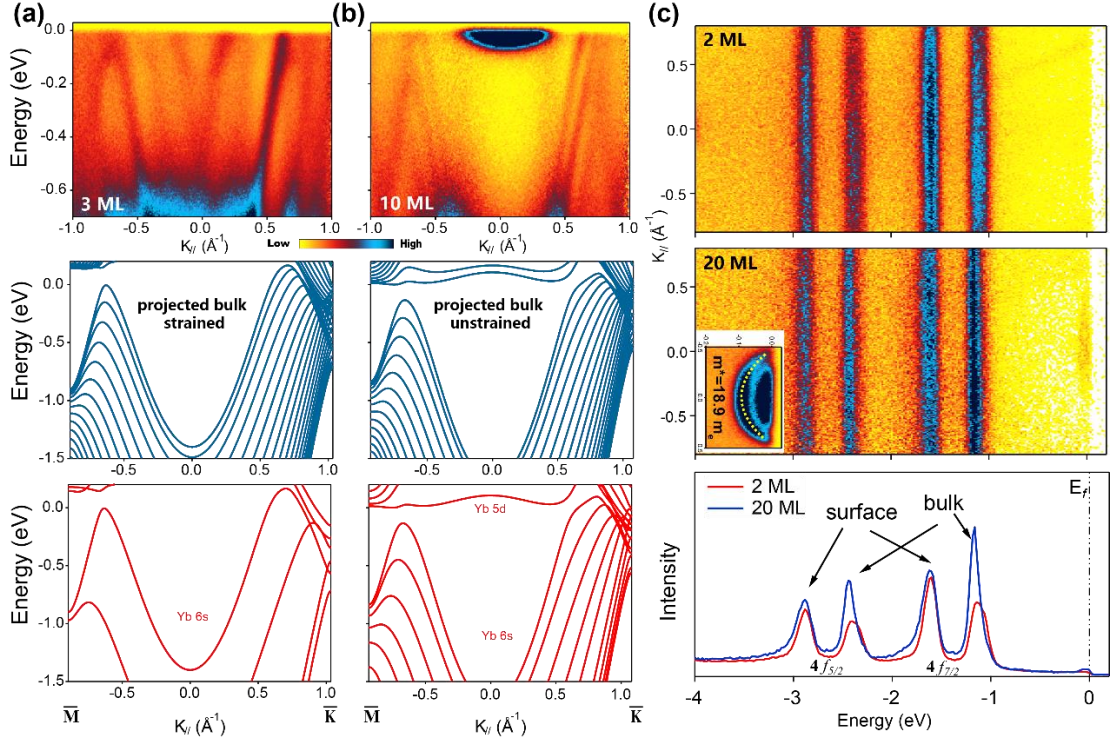

Fig. 2 (Color online). Thickness-dependent electronic structure of Yb films. (a,b). ARPES spectra (top panels) of the magic-height islands (3 ML coverage, a) and the bulk phase (10 ML, b), in comparison with DFT calculations (middle and bottom panels). The middle panels show the projected bulk band calculations for a 20-ML slab along two high-symmetry directions, using the experimentally determined lattice constant, i.e., ~5% tensile strained for (a) and bulk for (b). The bottom panels show the calculated QWSs based on Eq. (1), assuming a constant phase shift at the surface/interface. (c) ARPES spectra over a wide energy range and integrated energy distribution curves (EDCs) for the magic-height islands (2 ML coverage) and the bulk phase (20 ML). The inset is a zoom-in view of the band dispersion near the zone center for the 20 ML film, highlighting a heavy electron band with an effective mass ~19 $m_e$.

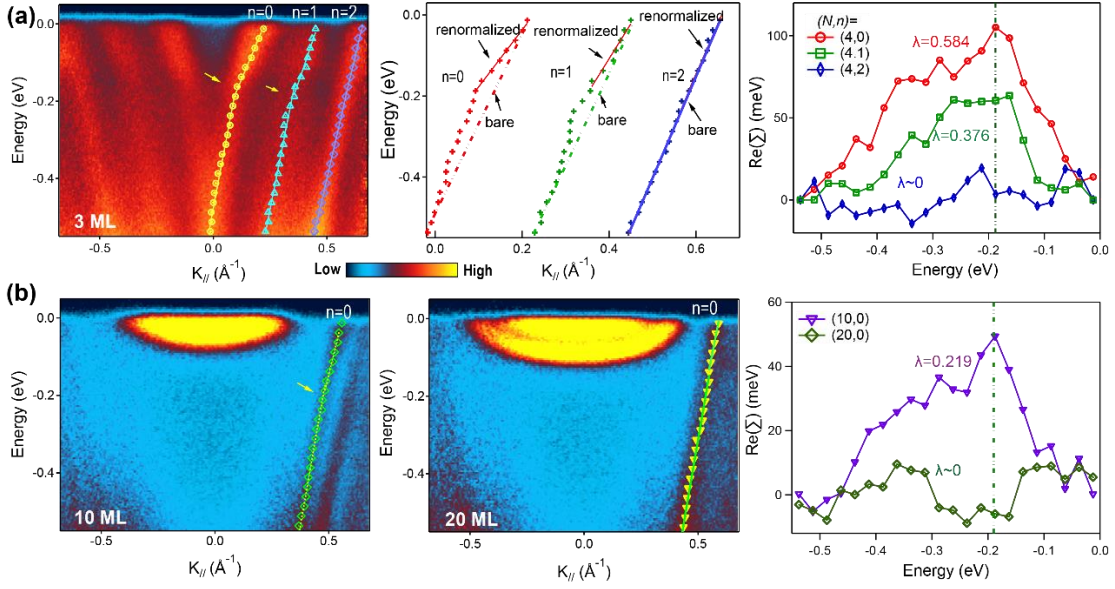

Fig. 3 (Color online). Pronounced kinks in QWSs of ultrathin Yb films by interfacial EPC. (a) Left: ARPES spectra (using He I photons) near the Fermi level for the magic-height islands (3 ML coverage). Strong kinks are seen at about -0.2 eV (yellow arrows). The extracted band dispersions based on an analysis of momentum distribution curves (MDCs) are overlaid on top. Middle: Extraction of the self-energy and EPC constant $\lambda$ based on the ARPES band dispersions (colored crosses). The dashed curves are the dispersions of the bare bands (without EPC). Right: Extracted real part of the self-energy and $\lambda$ for each QWS. Each QWS is labelled by the film thickness and quantum number ($N$, $n$). (b) Similar analysis for the bulk phase at 10 and 20 ML. Left and middle: ARPES spectra for 10 and 20 ML films, together with extracted band dispersions. Right: Extracted real part of the self-energy and $\lambda$ for (10,0) and (20,0) QWSs.

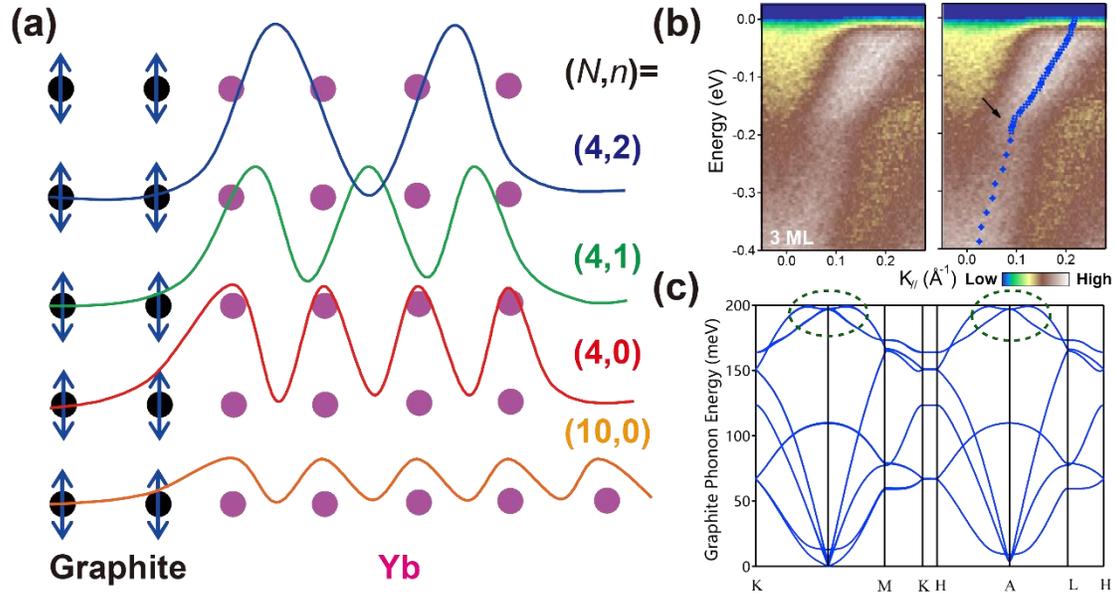

Fig. 4 (Color online). A simple model to explain the thickness (*N*) and subband (*n*) dependence of interfacial EPC constant $\lambda$. (a) Drawings of the simplified probability density of QWS wave functions, labelled by (*N*, *n*). Black (purple) filled circles indicate carbon (Yb) atoms, and blue arrows indicate movement directions of carbon atoms for the phonons involved in interfacial EPC. The probability density for QWSs are shown as curves, with their zeros offset vertically for clarity. $\lambda$ is approximately proportional to the square of the probability density at the interface. (b) A zoom-in view of the kink for the (4,0) QWS (data from 3 ML coverage), with its extracted dispersion and kink position shown on the right. (c) Calculated phonon dispersion relations for graphite, with dashed green circles highlighting the relevant phonon modes involved in the interfacial EPC.

16